\documentclass[reprint,pre,showpacs,aps]{revtex4-1}
\usepackage{amsmath}
\usepackage{graphicx}
\usepackage{bm}

\begin{document}

\title{Transfer matrix method for optics in graphene layers}
\author{Tianrong Zhan, Xi Shi, Yunyun Dai, Xiaohan Liu}
\author{Jian Zi}\email{jzi@fudan.edu.cn}
\affiliation{Department of Physics, Key Laboratory of Micro and Nano Photonic Structures
(MOE), and State Key Laboratory of Surface Physics, Fudan
University, Shanghai 200433, P. R. China}


\begin{abstract}
A transfer matrix method is developed for optical calculations of
non-interacting graphene layers. Within the framework of this method,
optical properties such as reflection, transmission and absorption for
single-, double- and multi-layer graphene are studied. We also apply
the method to structures consisting of periodically arranged graphene
layers, revealing well-defined photonic band structures and even photonic
bandgaps. Finally, we discuss graphene plasmons and introduce a
simple way to tune the plasmon dispersion.
\end{abstract}


\maketitle

\section{Introduction}
\label{intro}

Graphene is a flat monolayer of graphite with carbon atoms closely packed in
a two-dimensional honeycomb lattice. A hallmark of graphene is the existence
of Dirac cones in the electronic band structure, resulting in extraordinary
structural and electronic properties with great potential for
nanoelectronics \cite{nov:04,wil:06,cas:09}.

In addition to outstanding electrical, mechanical and chemical properties,
graphene has interesting optical response. One of the striking optical
properties of graphene is that its reflectance, transmittance and absorbance
are determined by the fine structure constant \cite{nai:08,kuz:08}. Despite being
only one-atomic-layer thick with negligible reflection, a single
free-standing graphene shows significant absorbance, universally
about $2.2\%$ in a spectral range from near-infrared to visible \cite
{nai:08,sta:08,mak:08}. In the infrared regime, graphene absorption can be
altered by applying gate voltages \cite{wan:08,li:08}. For few-layer graphene,
the optical absorption is proportional to the number of layers \cite{cas:07},
leading to a visual image contrast which can be used practically to identify
the number of graphene layers on a substrate. The highly transparent and
outstanding electrical properties of graphene make it attractive as transparent
electrodes \cite{bae:10,jo:10}. The broadband absorption implies that
graphene has the potential as an active medium used in broadband
photodetectors \cite{xia:09,sta:12}, ultra-fast lasers \cite{sun:10} and optical
modulations \cite{liu:11}.

In doped or gated graphene, collective
excitations---plasmons exist with interesting optical features such as deep
subwavelength and high confinement of optical fields \cite
{wun:06,hwa:07,mik:07,liu:08,jab:09,pro:10,fei:11}, similar to surface
plasmons in metal surfaces \cite{rae:98,bar:03,mai:07}. As a result, graphene
may serve as a one-atom-thick platform for infrared and terahertz
metamaterials \cite{ju:11,vak:11}. A number of photonic devices such as
waveguides, splitters and combiners and superlenses could be envisioned
\cite{vak:11, nik:11}. Numerical simulations suggest that periodically patterned
arrays of doped graphene nanodisks may completely absorb infrared light at
certain resonant wavelengths \cite{tho:12}, which was soon confirmed
experimentally \cite{yan:12}. These interesting optical properties of graphene
may also offer potential applications in photonics \cite{bon:10}.

In this paper, we develop a transfer matrix method for studying optical
properties in non-interacting graphene layers. This paper is organized as
follows. In section \ref{formu}, we introduce the transfer matrix method for
studying the propagation of light through graphene layers, together with the
optical conductivity of graphene used in our calculations. In Secs. \ref{rta}
to \ref{plas}, we apply the transfer matrix method to the study of optics in
graphene layers. Specifically, in section \ref{rta} we discuss reflection,
transmission and absorption in single-, double- and multi-layer graphene.
In section \ref{pbg}, we discuss photonic band structures in periodical
graphene layers. In section \ref{plas}, we discuss plasmons in graphene.
Finally, we present our summary in section \ref{summ}.

\section{General formulation}

\label{formu}

Transfer matrix method is a powerful tool in the analysis of light
propagations through layered dielectric media \cite{yeh:88,zi:98}. The
central idea lies that electric or magnetic fields in one position can be related
to those in other positions through a transfer matrix. Within the framework
of the transfer matrix method, there are two kinds of matrices: one is the
transmission matrix that connects the fields across an interface and the
other is the propagation matrix that connects the fields propagating over a
distance within a homogeneous medium.

\subsection{Transmission matrix}

\begin{figure}[t]
\centering\includegraphics[angle=0,width=8.2cm]{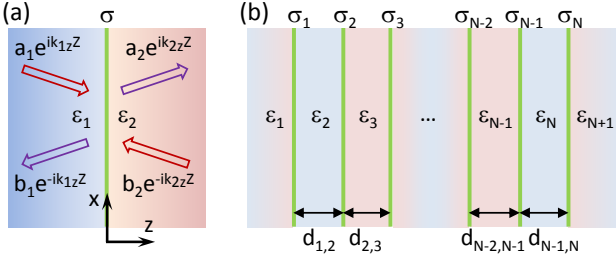}
\caption{ (a) A single graphene layer surrounded by two
dielectrics of dielectric constants $\varepsilon _{1}$ and
$\varepsilon _{2}$. The graphene layer is characterized by a
conductivity $\sigma $. Red and purple arrows indicate incoming and
outgoing light, respectively. (b) A stack of $N$ graphene layers of
conductivity $\sigma _{i}$ ($i=1,2,...,N$) are separated by
different dielectrics of dielectric constants $\varepsilon _{i}$
($i=1,2,...,N+1$). The spacing between two adjacent graphene layers is denoted
by $d_{i,i+1}$ ($i=1,2,...,N-1$).}
\label{fig1}
\end{figure}

We first consider the propagation of light across an interface formed by a
graphene layer which separates two dielectrics of dielectric constants
$\varepsilon _{1}$ and $\varepsilon _{2}$, as shown schematically in
figure~\ref{fig1}(a). The graphene layer has an optical conductivity $\sigma $
lying at $z=0$. Light is assumed to be polarized in the $y$ direction and
propagate in the $z$ direction. For structures considered,
$s$ and $p$ polarizations
can be decoupled. As a result, we can deal with $s$ and $p$ polarizations
separately.

For $p$ polarization, the magnetic field is polarized along the $y$
direction and can be written as the form
\begin{eqnarray}
H_{1y}&=& ( a_{1}e^{ik_{1z}z}+b_{1}e^{-ik_{1z}z}) e^{ik_{1x}x},
\qquad z<0, \\
H_{2y}&=& ( a_{2}e^{ik_{2z}z}+b_{2}e^{-ik_{2z}z}) e^{ik_{2x}x},
\qquad z>0.
\end{eqnarray}
Here, $a_{i}$ and $b_{i}$ $(i=1,2)$ are the field coefficients,
$k_{ix}$ ($k_{iz}$) is the $x$ ($z$) component of the wave-vector
$k_{i}=\sqrt{\varepsilon _{i}}\omega /c$,
where $\omega $ is the angular frequency and $c$ is the speed of light
in vacuum.
The first (second) term in the parenthesis on the right side represents
propagating waves along the $z$ ($-z$) direction. From the Snell's
law, we will immediately have $k_{1x}=k_{2x}$.

The electric and magnetic fields at the interface can be related by the
following boundary conditions \cite{jac:01}
\begin{eqnarray}
&\mathbf{n}\times (\mathbf{E}_{2}-\mathbf{E}_{1})|_{z=0} =0,  \label{bc-e} \\
&\mathbf{n}\times (\mathbf{H}_{2}-\mathbf{H}_{1})|_{z=0} =\mathbf{J},
\label{bc-h}
\end{eqnarray}
where $\mathbf{n}$ is the unit surface normal and $\mathbf{J}$ is the surface
current density of the graphene layer. Applying the above boundary
conditions at $z=0$, we will have
\begin{eqnarray}
&&\frac{k_{1z}}{\varepsilon _{1}}(a_{1}-b_{1})-\frac{k_{2z}}{\varepsilon _{2}
}(a_{2}-b_{2})=0,  \label{bc-p1} \\
&&(a_{1}+b_{1})-(a_{2}+b_{2}) =J_{x}.  \label{bc-p2}
\end{eqnarray}

Note that $\mathbf{J}$ can be obtained from Ohm's law, namely
\begin{equation}
J_{x}=\sigma E_{x}|_{z=0}=\frac{\sigma k_{2z}}{\varepsilon _{0}\varepsilon
_{2}\omega }(a_{2}-b_{2}),  \label{ohm-h}
\end{equation}
where $\varepsilon _{0}$ is the vacuum permittivity. Combing equations
(\ref{bc-p1})--(\ref{ohm-h}), the coefficients $a_{1}$ and $b_{1}$ can be
related to $a_{2}$ and $b_{2}$ by a 2$\times $2 transmission matrix
$D_{1\rightarrow 2}$
\begin{equation}
\left[
\begin{array}{c}
a_{1} \\
b_{1}%
\end{array}
\right] =D_{1\rightarrow 2}\left[
\begin{array}{c}
a_{2} \\
b_{2}%
\end{array}
\right] ,
\end{equation}
where
\begin{equation}
D_{1\rightarrow 2}=\frac{1}{2}\left[
\begin{array}{cc}
1+\eta _{p}+\xi _{p} & 1-\eta _{p}-\xi _{p} \\
1-\eta _{p}+\xi _{p} & 1+\eta _{p}-\xi _{p}%
\end{array}
\right] ,  \label{tran-p}
\end{equation}
with the parameters $\eta _{p}$ and $\xi _{p}$ given by
\begin{equation}
\eta _{p}=\frac{\varepsilon _{1}k_{2z}}{\varepsilon _{2}k_{1z}},\qquad \xi
_{p}=\frac{\sigma k_{2z}}{\varepsilon _{0}\varepsilon _{2}\omega }.
\end{equation}

For $s$ polarization, the electric field is polarized along the $y$
direction. Similarly, by applying the boundary conditions and Ohm's law,
the transmission matrix for $s$ polarization that relates the electric
fields at the two sides of the interface can be obtained, as
\begin{equation}
D_{1\rightarrow2}=\frac{1}{2}\left[
\begin{array}{cc}
1+\eta_{s}+\xi_{s} & 1-\eta_{s}+\xi_{s} \\
1-\eta_{s}-\xi_{s} & 1+\eta_{s}-\xi_{s}
\end{array}
\right] ,  \label{tran-s}
\end{equation}
with the parameters $\eta_{s}$ and $\xi_{s}$ given by
\begin{equation}
\eta_{s}=\frac{k_{2z}}{k_{1z}},\qquad\xi_{s}=\frac{\sigma\mu_{0}\omega}
{k_{1z}},
\end{equation}
where $\mu _{0}$ is the vacuum permeability.

The transmission matrix for $s$ and $p$ polarizations across an interface
has similar forms except for the sign of $\xi $ in the off-diagonal
elements. Introducing a polarization dependent parameter
$\varsigma _{m}$, the transmission
matrix for both polarizations can have an identical form
\begin{equation}
D_{1\rightarrow 2,m}=\frac{1}{2}\left[
\begin{array}{cc}
1+\eta _{m}+\xi _{m} & 1-\eta _{m}-\varsigma _{m}\xi _{m} \\
1-\eta _{m}+\varsigma _{m}\xi _{m} & 1+\eta _{m}-\xi _{m}
\end{array}
\right] ,  \label{tran}
\end{equation}
where $m=(s,p)$ and $\varsigma _{p}=1$ and
$\varsigma _{s}=-1$.

\subsection{Propagation matrix}

We now consider the propagation of light in a homogeneous medium. It
can be shown \cite{yeh:88} that the electric or magnetic fields at
$z+\Delta z$ can be related to those at the $z$ position by
a 2$\times $2 propagation matrix
\begin{equation}
P(\Delta z)=\left[
\begin{array}{cc}
e^{-ik_{z}\Delta z} & 0 \\
0 & e^{ik_{z}\Delta z}
\end{array}
\right] .
\end{equation}

\subsection{Transfer matrix for multi-layer graphene}

For a stack of $N$ graphene layers shown in figure \ref{fig1}(b), the transfer
matrix can be obtained by transmission matrices across different interfaces
and propagation matrices in different homogeneous dielectric media.
Denote the field coefficients on the left side of the leftmost
graphene layer by $a_{1}$ and $b_{1}$ and those on the right side of the
rightmost graphene layer by $a_{N+1}$ and $b_{N+1}$. The two sets of field
coefficients are then related by a 2$\times $2 transfer matrix $\mathcal{M}$,
namely
\begin{equation}
\left[
\begin{array}{c}
a_{1} \\ b_{1}
\end{array}
\right] =\mathcal{M}\left[
\begin{array}{c}
a_{N+1} \\ b_{N+1}
\end{array}
\right] ,
\end{equation}
with
\begin{equation}
\mathcal{M}=D_{1\rightarrow 2}P(d_{1,2})D_{2\rightarrow 3}P(d_{2,3})
\cdots
P(d_{N-1,N})D_{N\rightarrow N+1}.  \label{t-tm}
\end{equation}

\subsection{Optical spectrum calculations}

With the transfer matrix, we can easily calculate the optical spectra such
as reflection, transmission and absorption for multi-layer graphene.
Suppose that light is incident from left upon the multi-layer graphene with
the reflection and transmission coefficients denoted respectively by $r$ and
$t$. It can be shown that these coefficients are given by the elements of
$\mathcal{M}$
\begin{eqnarray}
r &=&\frac{M_{21}}{M_{11}},  \label{r-co} \\
t &=&\frac{1}{M_{11}}.
\end{eqnarray}
And reflectance and transmittance can be calculated for both $s$ and $p$
polarizations as
\begin{eqnarray}
R_{s,p}& =&|r_{s,p}|^{2},  \label{refl} \\
T_{s,p}& =&\eta _{s,p}|t_{s,p}|^{2},  \label{trans}
\end{eqnarray}
where
\begin{equation}
\eta _{s}=k_{(N+1)z}/k_{1z}, \qquad
\eta _{p}=\varepsilon_{1}k_{(N+1)z}/\varepsilon _{N+1}k_{1z}.
\end{equation}
Absorbance can then be readily obtained from
\begin{equation}
A=1-R-T.  \label{abso}
\end{equation}

\subsection{Optical conductivity of graphene}

For illustration and simplicity, in the present work we only consider the
situation where the chemical potential of graphene $\mu $ is much larger
than the temperature. In this situation, within the random-phase approximation
the optical conductivity of graphene $\sigma (\omega )$ is given
by \cite{wun:06,mik:07,hwa:07,and:02,gus:06,fal:07a}
\begin{equation}
\frac{\sigma (\Omega )}{\varepsilon _{0}c}=4\alpha \frac{i}{\Omega }+\pi
\alpha \left[ \vartheta (\Omega -2)+\frac{i}{\pi }\ln \left\vert \frac{
\Omega -2}{\Omega +2}\right\vert \right] .
\end{equation}
Here, $\Omega \equiv \hbar \omega /\mu $ is the dimensionless frequency,
$\alpha \equiv e^{2}/4\pi \varepsilon _{0}\hbar c$ ($\sim$1/137) is the fine
structure constant and $\vartheta (x)$ is the Heaviside step function. The
first and second terms on the right side stem from the intraband and
interband contributions, respectively.

\section{Reflection, transmission and absorption}

\label{rta}

\subsection{Single-layer graphene}

For a single graphene layer surrounded by two dielectrics of dielectric
constants $\varepsilon _{1}$ and $\varepsilon _{2}$, suppose that light is
incident from the dielectric medium of $\varepsilon _{1}$. The transfer
matrix is no other than the transmission matrix across the interface, given
by equations (\ref{tran-p}) and (\ref{tran-s}) for $p$ and $s$ polarizations,
respectively. From equations (\ref{refl}) and (\ref{trans}), the reflectance and
transmittance can be obtained as
\begin{eqnarray}
R_{s}& =&\left\vert \frac{\sqrt{\varepsilon _{1}}\cos \theta _{1}-\sqrt{
\varepsilon _{2}}\cos \theta _{2}-\widetilde{\sigma }}{\sqrt{\varepsilon _{1}
}\cos \theta _{1}+\sqrt{\varepsilon _{2}}\cos \theta _{2}+\widetilde{\sigma }
}\right\vert ^{2}, \\
R_{p}& =&\left\vert \frac{\sqrt{\varepsilon _{2}}/\cos \theta _{2}-\sqrt{
\varepsilon _{1}}/\cos \theta _{1}+\widetilde{\sigma }}{\sqrt{\varepsilon
_{2}}/\cos \theta _{2}+\sqrt{\varepsilon _{1}}/\cos \theta _{1}+\widetilde{
\sigma }}\right\vert ^{2}, \\
T_{s}& =&\frac{4\sqrt{\varepsilon _{1}\varepsilon _{2}}\cos \theta _{1}\cos
\theta _{2}}{\left\vert \sqrt{\varepsilon _{1}}\cos \theta _{1}+\sqrt{
\varepsilon _{2}}\cos \theta _{2}+\widetilde{\sigma }\right\vert ^{2}}, \\
T_{p}& =&\frac{4\sqrt{\varepsilon _{1}\varepsilon _{2}}/\left( \cos \theta
_{1}\cos \theta _{2}\right) }{\left\vert \sqrt{\varepsilon _{2}}/\cos \theta
_{2}+\sqrt{\varepsilon _{1}}/\cos \theta _{1}+\widetilde{\sigma }\right\vert
^{2}},
\end{eqnarray}
where $\theta _{1}$ and $\theta _{2}$ are incident and refracted angles,
respectively and $\widetilde{\sigma }=\sigma /\varepsilon _{0}c$.
By neglecting the higher-order terms of $\widetilde{\sigma }$, the absorbance
is given by
\begin{eqnarray}
A_{s}& =&\frac{4\sqrt{\varepsilon _{1}}\cos \theta _{1}{\rm Re}(\widetilde{
\sigma })}{\left( \sqrt{\varepsilon _{1}}\cos \theta _{1}+\sqrt{\varepsilon
_{2}}\cos \theta _{2}+\widetilde{\sigma }\right) ^{2}}, \\
A_{p}& =&\frac{4\sqrt{\varepsilon _{1}}{\rm Re}(\widetilde{\sigma })/\cos
\theta _{1}}{\left( \sqrt{\varepsilon _{2}}/\cos \theta _{2}+\sqrt{
\varepsilon _{1}}/\cos \theta _{1}+\widetilde{\sigma }\right) ^{2}}.
\end{eqnarray}
Obviously, for a single free-standing graphene layer under normal incidence
its absorbance for $\Omega >2$ is given by $\pi \alpha /(1+\pi \alpha
/2)^{2}\sim \pi \alpha$.

\begin{figure}[t]
\centering\includegraphics[angle=0,width=8.4cm]{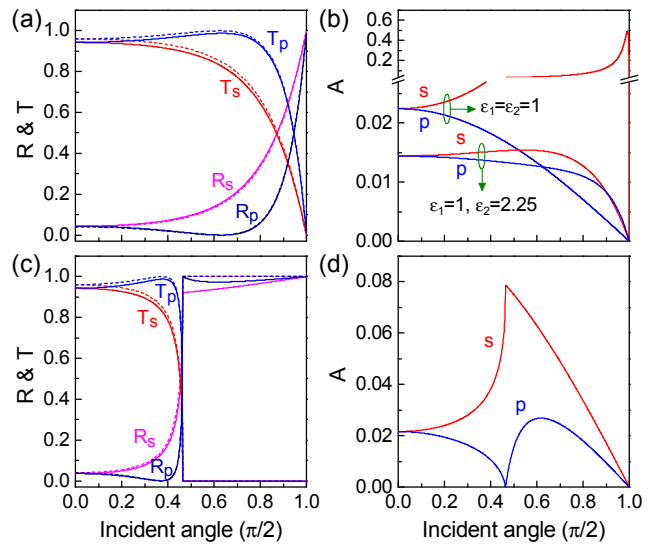}
\caption{ Reflectance, transmittance and absorbance of
single-layer graphene as a function of the incident angle for $\Omega >2$.
(a) Reflectance and transmittance for $s$ and $p$ polarizations with
$\varepsilon _{1}=1$ and $\varepsilon _{2}=2.25$. Dashed
lines represent those in the absence of the graphene layer. (b) Absorbance
for (i) $\varepsilon _{1}=\varepsilon _{2}=1$ and (ii) $\varepsilon _{1}=1$
and $\varepsilon _{2}=2.25$. (c) Same as (a) but for $\varepsilon _{1}=2.25$
and $\varepsilon _{2}=1$. The corresponding absorbance is given in (d).}
\label{fig2}
\end{figure}

In figure \ref{fig2}, the reflectance, transmittance and absorbance at
different incident angles for single graphene are shown. Light is
incident from the dielectric of $\varepsilon _{1}$. Reflection and transmission
are altered somewhat with respect to the case without the graphene layer.
For $\varepsilon _{1}\leq\varepsilon _{2}$, the absorbance decreases
monotonically with increasing incident angle for $p$ polarization, while for $s$
polarization it increases up to a maximum and then decreases monotonically.
For $\varepsilon _{1}=\varepsilon _{2}$, the absorbance for $s$ polarization
takes a universal value at normal incidence, about $\pi \alpha $ and the
maximal absorbance is 0.5 at an incident angle very close to $\pi /2$.

For $\varepsilon _{1}>\varepsilon _{2}$, total internal reflection is expected.
With the presence of the graphene layer, there is nearly no change
in the critical angle. However, above the critical angle reflectance is no
longer \textit{total} (smaller than one).
For $s$ polarization, the absorbance around the critical angle is several
times larger than the universal value of $\pi \alpha $. This implies
that the configuration of total internal
reflection could be exploited in measurements of optical conductivity of
graphene since it can suppress the signal-to-noise ratio considerably.

\subsection{Double-layer graphene}

Consider two graphene layers which are separated by a dielectric of
$\varepsilon_{2}$ with a thickness of $d$. The dielectric constant of the
leftmost dielectric is $\varepsilon _{1}$ and that of the rightmost dielectric
is $\varepsilon _{3}$. The transfer matrix of the structure can be
obtained from equation (\ref{t-tm}) as
\begin{equation}
\mathcal{M}=D_{1\rightarrow 2}P(d)D_{2\rightarrow 3},
\end{equation}
whose elements are given by
\begin{equation}
M_{\mu \nu ,m}=\frac{\cos k_{2z}d}{2}A_{\mu \nu ,m}+\frac{i\sin k_{2z}d}{2}
B_{\mu \nu ,m},
\end{equation}
where $\mu=1,2$ and $\nu=1,2$. The parameters $A_{\mu \nu ,m}$
and $B_{\mu \nu ,m}$ are given by
\begin{eqnarray}
A_{\mu \nu ,m}& =&1+(-1)^{\mu +\nu }\eta _{m}\eta _{m}^{\prime } \nonumber \\
&& -(-1)^{\nu }\left( \xi _{m}^{\prime }+\varsigma _{m}\xi _{m}\eta
_{m}^{\prime }\right) , \\
B_{\mu \nu ,m}& =&(-1)^{\mu }\eta _{m}+(-1)^{\nu }\left( \eta _{m}^{\prime
}+\varsigma _{m}\xi _{m}\eta _{m}^{\prime }\right) \nonumber \\
&& -(-1)^{\mu +\nu }\eta _{m}\xi _{m}^{\prime }-\varsigma _{m}\xi _{m},
\end{eqnarray}
where
\begin{equation}
\begin{array}{ll}
\eta _{s}=k_{2z}/k_{1z}, &
\eta _{p}=\varepsilon_{1}k_{2z}/\varepsilon _{2}k_{1z}, \\
\xi _{s}=\sigma \mu _{0}\omega /k_{1z}, &
\xi _{p}=\sigma k_{2z}/\varepsilon _{0}\varepsilon _{2}\omega , \\
\eta _{s}^{\prime }=k_{3z}/k_{2z}, &
\eta _{p}^{\prime }=\varepsilon_{2}k_{3z}/\varepsilon _{3}k_{2z}, \\
\xi _{s}^{\prime }=\sigma \mu _{0}\omega /k_{2z}, &
\xi _{p}^{\prime}=\sigma k_{3z}/\varepsilon _{0}\varepsilon _{3}\omega.
\end{array}
\end{equation}

\begin{figure}[t]
\centering\includegraphics[angle=0,width=8.4cm]{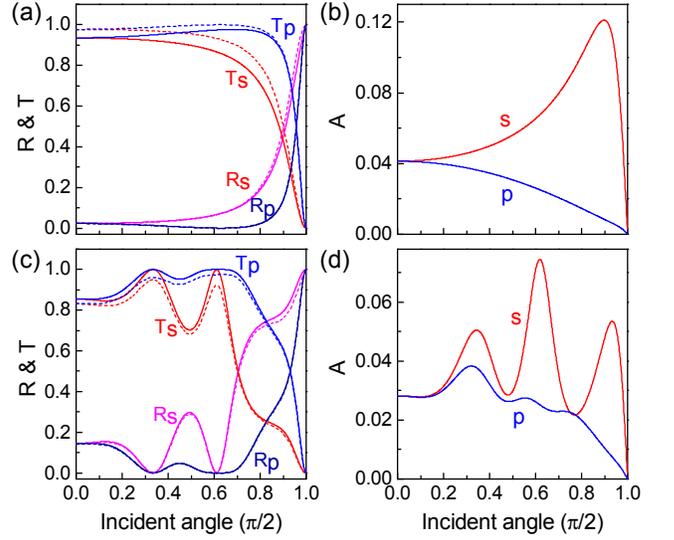}
\caption{ Reflectance, transmittance and absorbance of
double-layer graphene as a function of the incident angle for $\Omega >2$.
Two graphene layers have the same optical conductivity. (a) Reflectance and
transmittance for $s$ and $p$ polarizations with $d=0.1\hbar c/\mu $
(corresponding to 0.132 $\mu$m for a typical chemical potential $\mu=0.15$ eV).
Dashed lines represent those in the absence of the graphene layers. The
corresponding absorbance is given in (b). (c) Same as (a) but for $d=8\hbar
c/\mu $ (corresponding to 10.5 $\mu$m for $\mu=0.15$ eV).
The corresponding absorbance is given in (d). }
\label{fig3}
\end{figure}

In figure~\ref{fig3}, the reflectance, transmittance and absorbance of
double-layer graphene for $\varepsilon _{1}=\varepsilon _{3}=1$ and
$\varepsilon _{2}=2.25$ are shown. For small $d$, despite different values,
the dependence of the reflectance, transmittance and absorbance on the
incident angle is, in general, similar to that of single-layer graphene.
At normal incidence, the absorbance is nearly twice
the universal value of $\pi \alpha$. For large $d$,
however, there are oscillations in reflectance, transmittance and absorbance,
which originate from the thin-film interference.

\subsection{Multi-layer graphene}

For multi-layer graphene, the transfer matrix can be obtained from equation
(\ref{t-tm}). Reflectance, transmittance and absorbance can be then obtained from equations
(\ref{refl}), (\ref{trans}) and (\ref{abso}) respectively.

\begin{figure}[t]
\centering\includegraphics[angle=0,width=8.cm]{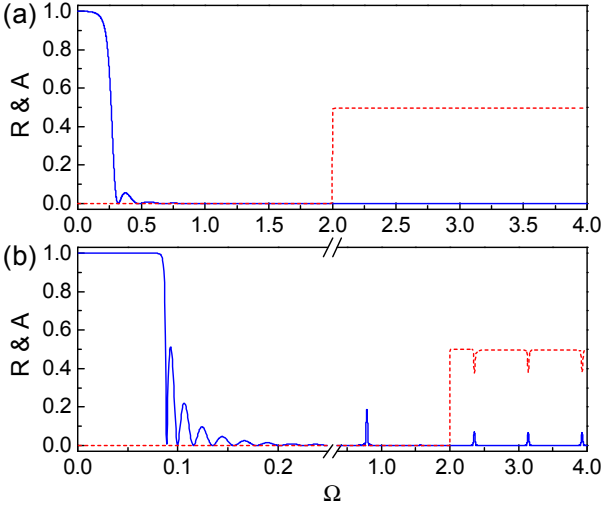}
\caption{ Reflectance (solid line) and absorbance (dashed
line) spectra of multi-layer graphene at normal incidence. There are totally
30 identical graphene layers. (a) $d=0.1\hbar c/\mu $ and (b)
$d=4\hbar c/\mu $ (corresponding to 0.132 and 5.26 $\mu$m for
$\mu=0.15$ eV, respectively). }
\label{fig4}
\end{figure}

In figure~\ref{fig4}, the reflectance and absorbance of multi-layer graphene
at normal incidence are shown. The structure consists of identical graphene
layers in air, separated equally by a distance $d$. At low frequencies the
reflectance is close to one up to a certain cutoff frequency. Above the
cutoff frequency, for small $d$ the reflectance oscillates and drops rapidly to
zero. For large $d$, however, there appear sharp reflection peaks for
frequency above the cutoff frequency, resulting from the multiple
interference by the graphene layers.

The structure has zero absorbance for $\Omega <2$
and shows remarkable absorption for $\Omega >2$. For small $d$, the
absorbance is nearly a constant, being 0.5 for $\Omega >2$. With increasing
$d$, multiple interference by graphene layers may play an important role,
leading to sharp absorption dips. These interesting properties imply that
multi-layer graphene has the potential as dark materials for achieving lower
reflection coatings and enhanced photodetection \cite{lud:11}.

\section{Photonic band structure of periodical graphene layers}

\label{pbg}

When dielectrics are arranged in a periodical way to form so-called photonic
crystals \cite{yab:87,joh:87,joa:08}, electromagnetic waves should be
strongly modulated by Bragg scatterings, showing photonic
band structures with well-defined photonic bands and even photonic bandgaps.
For graphene layers stacking in a periodical way, photonic band structures
should also be expected due to the introduced periodical modulations.

\begin{figure}[t]
\centering\includegraphics[angle=0,width=8.4cm]{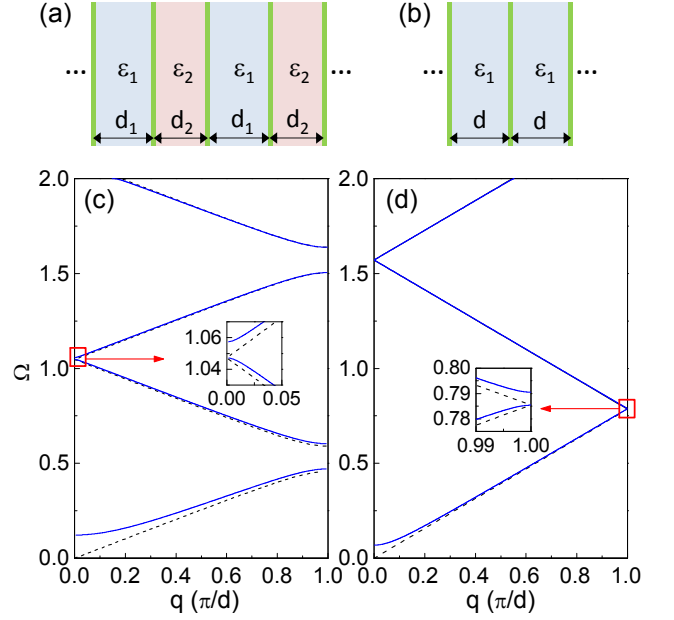}
\caption{ (a) Schematic of a structure consisting of
identical graphene layers placed at the interfaces of a one-dimensional
photonic crystal. (b)
Schematic of a structure consisting of identical graphene layers separated
equally in air. (c) Photonic band structure (solid lines) for the structure
in (a). The parameters are $\varepsilon _{1}=1$, $\protect%
\varepsilon _{2}=2.25$, $d_{1}=3\hbar c/\mu $ and $d_{2}=2\hbar c/
\mu $ (corresponding to 2.63 and 3.95 $\mu$m for $\mu=0.1$ eV,
respectively). Dashed lines are the results in the absence of graphene
layers. (d) Same as (c) but for the structure in (b). The parameters are
$\varepsilon =1$ and $d=4\hbar c/\mu $. Dashed lines are
simply the folded dispersion in air $\omega =qc$.}
\label{fig5}
\end{figure}

The first structure considered is shown schematically in figure
\ref{fig5}(a). It is composed of identical graphene layers embedded into the
interfaces of a one-dimensional photonic crystal consisting of two
dielectrics of $\varepsilon _{1}$ and $\varepsilon _{2}$ with a thickness of
$d_{1}$ and $d_{2}$, respectively. For such a structure, its transfer matrix
after propagating over one unit cell reads
\begin{equation}
\mathcal{M}_{m}=D_{1\rightarrow 2,m}P(d_{2})D_{2\rightarrow 1,m}P(d_{1}).
\end{equation}
The photonic band structure can be then obtained from the diagonal elements
of the transfer matrix \cite{zi:98},
$\cos (qd)=\left( M_{11,m}+M_{22,m}\right) /2$,
where $q$ is the Bloch wave-vector and $d=d_{1}+d_{2}$ is the lattice
constant. It can be explicitly written as
\begin{eqnarray}
\cos (qd)&=&\cos (k_{1z}d_{1})\cos (k_{2z}d_{2})- \frac{1}{2}(\eta _{m}+\eta
_{m}^{-1})  \nonumber \\
&&\times \sin (k_{1z}d_{1})\sin (k_{2z}d_{2})- \Delta _{m},  \label{band}
\end{eqnarray}
where
\begin{eqnarray}
\Delta _{m}& =& i\xi _{m}[\left( 1+\eta _{m}^{-1}\right) \sin
(k_{1z}d_{1}+k_{2z}d_{2})  \nonumber \\
&& +\varsigma _{m}\left( 1-\eta _{m}^{-1}\right) \sin
(k_{1z}d_{1}-k_{2z}d_{2})]  \nonumber \\
&& +\frac{\xi _{m}^{2}}{2\eta _{m}}\sin (k_{1z}d_{1})\sin (k_{2z}d_{2}).
\end{eqnarray}
Without the term $\Delta _{m}$, equation (\ref{band}) reduces to the photonic
band structure of a one-dimensional photonic crystal \cite{zi:98}.

For identical graphene layers separated equally by a distance $d$ in air as
shown in figure \ref{fig5}(b), the transfer matrix simply reads
$\mathcal{M}_{m}=D_{m}P(d)$,
where
\begin{equation}
D_{m}=\left[
\begin{array}{cc}
1+\xi _{m}/2 & -\varsigma _{m}\xi _{m}/2 \\
\varsigma _{m}\xi _{m}/2 & 1-\xi _{m}/2
\end{array}
\right] .
\end{equation}
The photonic band structure of the structure is then given by
\begin{equation}
\cos (qd)=\cos (k_{z}d)-\frac{i\xi _{m}}{2}\sin (k_{z}d),
\end{equation}
which is identical to that given in reference \cite{fal:07}.

In figure~\ref{fig5}, the photonic band structures of periodical graphene
layers are shown for the propagation direction perpendicular to the graphene
layers. For both structures the first
photonic band starts from a certain nonzero cutoff frequency, different from
conventional dielectric photonic crystals. This cutoff frequency corresponds
exactly to that observed in the reflection spectra shown in figure \ref{fig4}.
Below the cutoff frequency, high reflection or low transmission is expected,
which was also observed in numerical simulations of a stack of graphene
layers separated by dielectric slabs \cite{kai:12}. It is known that in the
low-frequency limit periodical metallic structures can be considered as bulk
metals with a depressed effective plasma frequency \cite{pen:96, zi:05}.
Thus, periodical graphene layers can also be regarded as a bulk metal with
an extremely low effective plasma frequency.

For the structure shown in figure \ref{fig5}(a), the photonic crystal in
the absence of graphene layers displays well-defined photonic bands and
bandgaps. With the introduction of graphene layers, however, both photonic
bands and bandgaps are modified. For example, for the photonic crystal in the
absence of graphene layers, there should be no bandgap between the second
and third photonic bands since this photonic crystal is a quarter-wave stack. In
the presence of graphene layers, however, a mini photonic bandgap opens up.

For the second structure shown in figure \ref{fig5}(b), there should be no
photonic bandgaps in the absence of graphene layers. In the presence of
graphene layers, however, a series of mini photonic bandgaps appear owing to
the multiple interference by graphene layers. It is known that for frequency
within photonic bandgaps light propagation is forbidden \cite{yab:87,joh:87,joa:08}.
For a structure consisting of finite graphene layers, this will
cause strong reflection for frequency located into the mini
photonic bandgaps, as can be clearly seen from figure \ref{fig4}(b).

\section{Plasmons}

\label{plas}

A flat metal surface can support surface plasmons \cite{rae:98,bar:03,mai:07}
which are transverse magnetic (TM) electromagnetic
waves coupled with collective oscillations of surface charges. Surface
plasmons can propagate along the metal surface with the fields decaying
exponentially away from the both sides of the surface. In doped or gated
graphene, free carriers can also support
plasmons \cite{wun:06,hwa:07,mik:07,liu:08,jab:09,pro:10}. Owing to the
two-dimensional nature and
unique electronic band structure, graphene can support not only TM but also
transverse electric (TE) plasmons \cite{mik:07}. The later does no exist in
conventional metal surfaces.

For a graphene layer surrounded by two dielectrics as shown in
figure \ref{fig1}(a), the transfer matrix is simply the transmission matrix. From
equation (\ref{r-co}), the reflection coefficient of the system can be obtained.
The condition for the existence of plasmons
is that the reflection coefficient has poles, namely
\begin{equation}
1+\eta _{m}+\varsigma _{m}\xi _{m}=0,  \label{cond-pl}
\end{equation}
where the subscript $m$ stands for $s$ and $p$ polarization, corresponding to
TE and TM modes, respectively.

From equation (\ref{cond-pl}), the dispersion of TM plasmons can be obtained as
\begin{equation}
\frac{\varepsilon _{1}}{\sqrt{Q^{2}-\varepsilon _{1}\Omega ^{2}}}+\frac{
\varepsilon _{2}}{\sqrt{Q^{2}-\varepsilon _{2}\Omega ^{2}}}=-\frac{i\sigma
(\Omega )/\varepsilon _{0}c}{\Omega },
\end{equation}
with $Q\equiv \hbar cq/\mu $, where $q$ ($\equiv k_{x}$) is the wave-vector
of the plasmons.
Obviously, TM plasmons can exist if the imaginary part of
$\sigma $ is \textit{positive}. From the above equation, TM plasmons are far
below the light line, i.e., $q\gg \omega /c$. Thus, in this non-retarded
regime, the dispersion of TM plasmons is simplified to
\begin{equation}
Q=-\left( \varepsilon _{1}+\varepsilon _{2}\right) \Omega
(\varepsilon _{0}c)/i \sigma.
\label{tm-pl}
\end{equation}
For $\Omega >2$, $\sigma $ has a real value as well, leading to a strong
loss due to interband excitations. For small $q$, the dispersion of TM
plasmons reduces to
\begin{equation}
\Omega =2\sqrt{\frac{\alpha }{\left( \varepsilon _{1}+\varepsilon
_{2}\right) }Q},
\end{equation}
which shows the known $\sqrt{q}$-dependence \cite{wun:06,hwa:07}.

From equation (\ref{cond-pl}), the dispersion of TE plasmons is given by
\begin{equation}
\sqrt{Q^{2}-\varepsilon _{1}\Omega ^{2}}+\sqrt{Q^{2}-\varepsilon _{2}\Omega
^{2}}=\frac{i\sigma }{\varepsilon _{0}c}\Omega ,
\end{equation}
which is the same as that given in reference \cite{mik:07}. TE plasmons can exist
if $\varepsilon _{1}=\varepsilon _{2}$ and the imaginary part of $\sigma $
is \textit{negative} (for $\Omega >1.667$). Note that the term on the right
side of the above equation is very small in the frequency window
$1.667<\Omega <2$. As a result, the dispersion of TE plasmons should be below but very
close to the light line $\omega =qc/\sqrt{\varepsilon _{1}}$.

\begin{figure}[t]
\centering\includegraphics[angle=0,width=8.4cm]{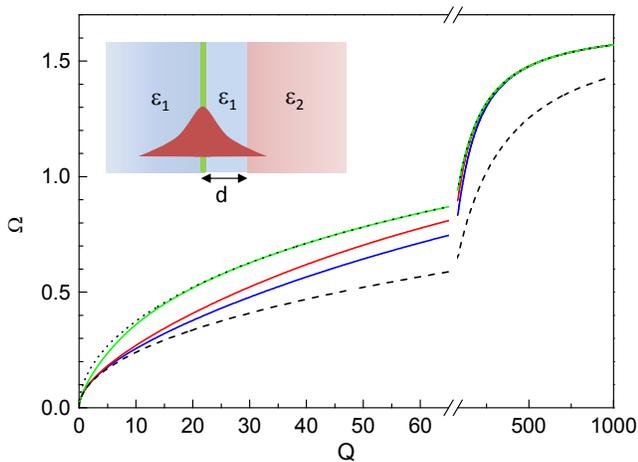}
\caption{ Dispersion of TM plasmons for a structure shown in
the inset where the graphene layer is apart from a dielectric substrate of
$\varepsilon _{2}=10$ over a distance of $d$ and the other
dielectrics are air with $\varepsilon _{1}=1$. Blue, green and red
lines are the results for $\tilde{d}=0.005$, 0.01 and 0.1, respectively. The
dotted line represents the
dispersion for a free-standing graphene layer in air and the dashed line
corresponds to the case $d=0$.}
\label{fig6}
\end{figure}

We now consider a structure where a graphene layer is apart from a
dielectric substrate of $\varepsilon _{2}$ over a distance of $d$, as shown
schematically in the inset of figure \ref{fig6}. For the structure, TE
plasmons do not exist for finite $d$ and thus we only discuss TM plasmons.
For $p$ polarization, the transfer matrix of the structure can be obtained
from equation (\ref{t-tm}), namely
\begin{equation}
\mathcal{M}=DP(d)D^{\prime },
\end{equation}
where
\begin{eqnarray}
D &=&\frac{1}{2}\left[
\begin{array}{cc}
1+\eta _{p}+\xi _{p} & 1-\eta _{p}-\xi _{p} \\
1-\eta _{p}+\xi _{p} & 1+\eta _{p}-\xi _{p}
\end{array}
\right] , \\
D^{\prime } &=&\frac{1}{2}\left[
\begin{array}{cc}
1+\eta _{p}^{\prime } & 1-\eta _{p}^{\prime } \\
1-\eta _{p}^{\prime } & 1+\eta _{p}^{\prime }
\end{array}
\right] ,\\
P&=&\left[
\begin{array}{cc}
e^{-ik_{1z}d} & 0 \\
0 & e^{ik_{1z}d}
\end{array}
\right] ,
\end{eqnarray}
with
\begin{equation}
\eta _{p}=1,\ \ \ \xi _{p}=\sigma k_{1z}/\varepsilon _{0}\varepsilon
_{1}\omega ,\ \ \ \eta _{p}^{\prime }=\varepsilon _{1}k_{2z}/\varepsilon
_{2}k_{1z}.
\end{equation}%
With the transfer matrix, it is easy to obtain the reflection coefficient
from equation (\ref{r-co}). The condition for the existence of plasmons thus
reads
\begin{eqnarray}
\left( 1+\eta _{p}+\xi _{p}\right)& &\left( 1+\eta _{p}^{\prime }\right)
e^{-ik_{1z}d}+  \nonumber \\
&&\left( 1-\eta _{p}-\xi _{p}\right) \left( 1-\eta _{p}^{\prime }\right)
e^{ik_{1z}d} =0.
\end{eqnarray}
Since the dispersion of graphene plasmons lies far below the light line, the
non-retarded condition $q\gg \omega /c$ still holds, leading to $\eta
_{p}^{\prime }\simeq \varepsilon _{1}/\varepsilon _{2}$ and $\xi _{p}\simeq
i\sigma q/\varepsilon _{0}\varepsilon _{1}\omega $. Thus, plasmon dispersion
can be simplified as
\begin{equation}
\frac{2\varepsilon _{1}\left( \varepsilon _{1}+\varepsilon _{2}\right) }
{\left( \varepsilon_{1}+\varepsilon _{2}\right) +\left( \varepsilon _{1}-\varepsilon
_{2}\right) e^{-2Q\tilde{d}}}=-\frac{i\sigma}{\varepsilon _{0} c}
\frac{Q}{\Omega },
\end{equation}
where $\tilde{d}=d\mu /\hbar c$.

In figure \ref{fig6}, the dispersion of TM plasmons for the structure shown
in the inset is given. The separation of the graphene layer from the dielectric
substrate of $\varepsilon _{2}$ affects considerably the dispersion. For
small $q$, the dispersion takes that for the case of $d=0$. For large $q$,
it approaches that for the free-standing case. This dispersion interchange
with increasing $q$ can be understood by the fact that the fields of
plasmons decay exponentially into the surrounding media, as schematically
depicted in the inset. For small $q$, the decay length is much larger than $d$
such that the fields concentrate dominantly in the substrate. As a result,
the dispersion should take that for the case $d\sim 0$.
For large $q$, the fields decay very fast such that the decay length is much
smaller than $d$. In this situation, the fields could not sense the substrate
dielectric layer.

Obviously, the dispersion interchange can be tuned by $d$. The change
from one kind of dispersion to the other occurs faster for large $d$ than for
small $d$. Practically, we may adopt this dispersion interchange to tune the
dispersion of TM plasmons, or in other words, the refractive
index of the plasmons. From equation (\ref{tm-pl}), the corresponding refractive
index of TM plasmons for a graphene layer surrounded by two dielectrics of
$\varepsilon _{1}$ and $\varepsilon _{2}$ is given by
\begin{equation}
n_{p}\equiv qc/\omega =-\frac{\varepsilon _{1}+\varepsilon _{2}}{i\sigma
/\varepsilon _{0}c}.
\end{equation}
With the structure shown in the inset of figure \ref{fig6},
the refractive index of TM plasmons can thus be tuned
from $-2\varepsilon_{1}/\left( i\sigma /\varepsilon _{0}c\right) $ to
$-\left( \varepsilon_{1}+\varepsilon _{2}\right) /\left( i\sigma /\varepsilon _{0}c\right) $.
This offers practically a simple approach to manipulate the dispersion of TM plasmon or the
refractive index by changing $d$.

\section{Conclusions}

\label{summ}

In the present paper, we developed a transfer matrix method for optical
calculations in non-interacting graphene layers. With the framework of this
method, the transfer matrix for various graphene layers can be obtained.
With the transfer matrix, reflectance, transmittance and absorbance spectra
of graphene layers can be easily obtained. In addition, photonic band
structures for periodical graphene layers and even graphene plasmons
can be studied in a rather simple way.

Using the transfer matrix method, we studied the optical properties such as
reflection, transmission and absorption for single-, double- and
multi-layer graphene. We showed that the configuration of total internal
reflection in single-layer graphene and thin-film interference effects in
double-layer graphene could be exploited to enhance the light absorption. For
multi-layer graphene, there exists a cutoff frequency below which the
reflectance is as high as one. For a small spacing distance, the absorbance
is as high as 50\% for $\Omega >2$. With increasing spacing distance,
sharp reflection peaks and absorption dips appear owing to the multiple
interference by graphene layers.

We applied the transfer matrix method to structures consisting of
periodically arranged identical graphene layers. The structures are
characterized by photonic band structures with well-defined photonic bands
and bandgaps. We revealed that these structures can be regarded as bulk
metals with an extremely low effective plasma frequency.

Finally, we discussed plasmons in a graphene layer which is apart from a
dielectric substrate. We found that the plasmon dispersion can be tuned by
the separating distance between the graphene layer and the dielectric
substrate. Our results show that the transfer matrix method could serve as a
versatile tool to study optical properties in graphene layers.

\section{Acknowledgment}
This work was supported by the 973 Program (Grant Nos. 2013CB632701 and
2011CB922004). The research of J.Z. and X.H.L. is further supported by the NSFC.


\end{document}